\documentstyle[12pt,aps]{revtex}

\def\lp {\left( }
\def\rp {\right) }
\def\lb {\left[ }
\def\rb {\right] }
\def\lc {\left\{ }
\def\rc {\right\} }
\def\ra {\rangle }
\def\la {\langle }
\def\rar {\rightarrow }

\def\beq{\begin{equation}}
\def\eeq{\end{equation}}
\def\bea{\begin{eqnarray}}
\def\eea{\end{eqnarray}}
\def\bb {\bibitem}
\def\nn {\nonumber}
\def\ni {\noindent}
\def\vs {\vspace}

\def\di {\partial_\mu }
\def\ds {\partial^\mu }

\def\Vb {\bar{V}}

\def\a{\alpha}
\def\b{\beta}
\def\d {\delta}
\def\e{\epsilon}
\def\g{\gamma}
\def\D {\Delta}

\def\l{\lambda}
\def\L{\Lambda}
\def\m{\mu}

\def\o{\omega}
\def\p {\pi}
\def\r{\rho}
\def\s{\sigma}

\def\hm {\hat{m}}
\def\qb {\bar{q}}

\def\bfi {\mbox{\boldmath $\phi$}}
\def\bk {\mbox{\boldmath $k$}}
\def\bq{\mbox{\boldmath $q$}}
\def\br {\mbox{\boldmath $r$}}

\def\bsg {\mbox{\boldmath $\sigma$}}
\def\btu {\mbox{\boldmath $\tau$}}

\def\bz{\mbox{\boldmath $0$}}

\begin{document}

\title{\Large \bf  QUARK CONDENSATE IN THE DEUTERON}
\author{J-L. Ballot}
\address{Division de Physique Th\'eorique, Institut de Physique
Nucl\'eaire\\ F-91406, Orsay CEDEX, France}

\author{M. Ericson}
\address{Institut de Physique Nucl\'eaire et IN2P3,CNRS, Universit\'e Claude Bernard Lyon I\\
43 Bd. du 11 Novembre, F-69622, Villeurbanne CEDEX, France\\
and\\
CERN, CH 1211, Geneva 23, Switzerland}

\author{ M.R. Robilotta }
\address{FINPE, Instituto de F\'{\i}sica, Universidade de S\~ao Paulo \\
C.P. 66318, 05389-970, S\~ao Paulo, SP, Brasil}
\date{\today}

\maketitle

abstract:

We study the changes produced by the deuteron on the QCD quark condensate by means the Feynman-Hellmann theorem and
find that the pion mass dependence of the pion-nucleon coupling could play an important role.
We also discuss the relation between the many body effect of the condensate and the meson exchange currents, as seen by photons and pions. 
For pion probes, the many-body term in the physical amplitude differs significantly from that of soft pions, the one linked to 
the condensate. Thus no information about the many-body term of the condensate can be extracted from the pion-deuteron
scattering length.
On the other hand, in the Compton amplitude, the relationship with the condensate is a more direct one.

\section{introduction}

The QCD vacuum has a complex structure, with condensates of quarks and gluons, that can be disturbed by the presence of hadronic matter. 
In the case of nucleons, for instance, valence quarks give rise to an anti-screening interaction, which reduces the magnitude of the condensate.
This gives rise to the nucleon sigma-term ($\s_N$), that can be extracted from pion-nucleon scattering.

In the case of  nuclei, in first approximation the effects of idependent nucleons add up \cite{CFG92}. 
But as  nucleons are interacting, there also exist modifications of the condensate due to the nucleon-nucleon potential. 
It is reasonable to believe that the influence of this potential is more  important in large nuclei, but the study of these systems is complicated and requires simplifying approximations. 
Therefore it is interesting to look for effects of the NN interaction over the condensate in light nuclei. 
The deuteron, in particular, has been extensively explored and allows calculations with little theoretical uncertainties.

In principle, one should use QCD to study the reaction of the quark condensate to the presence of hadronic matter.
However, as this is beyond our present capabilities, we use effective interactions of colourless hadrons in place of the fundamental ones.
Effective theories should be as close as possible to QCD and, in particular, share its symmetries.
The interactions of quarks and gluons are approximately invariant under chiral transformations and broken, in the $SU(2)$
sector, by the very small quark masses.
Therefore, at the hadron level, one requires the effective theory to possess approximate chiral symmetry, now
broken by $\mu$, the pion mass.

In the case of NN interactions, most of the dynamics relevant at large and intermediate energies can be described, in the framework of effective theories,  by exchanges of one and two pions \cite{P80,B87,Ch}. 
For the short distance region, on the other hand, neither meson nor quark models produce precise quantitative predictions and realistic potentials must rely on free parameters.
In the case of the deuteron, these short distance uncertainties are minimized,  for it is heavily dominated by the one pion exchange potential (OPEP) \cite{TERC,BER,BR}.

In this work we discuss the disturbances of the QCD vacuum produced by the deuteron.
In sect. II,  we concentrate on the dependence of its binding energy on the quark mass, to derive the quark 
condensate using the Feynman-Hellmann theorem.
The changes induced in the quark condensate by the nuclear force can be related to exchange currents, as probed by means of both photons and pions. 
Thus, in sect. III we discuss the case of electromagnetic probes and in sect. IV we study $\pi$d scattering.
Finally, in sect. V we present our results and discuss how they are related to measurable quantities.


\section{Feynman-Hellmann}

The deuteron mass is written as $ M = 2 m - \e $, where m is the nucleon mass and $ \e $ is the binding energy, which we take as positive. 
The part of $M$ due to chiral symmetry breaking corresponds to the deuteron sigma-term, given by

\beq
\s_d = - \int d^3 r 
\lp\la d \right| {\cal L}_{SB} \left| d \ra
-\la 0 \right|  {\cal L}_{SB} \left| 0 \ra \rp  \;,
\label{B1}
\eeq

\ni
where $ {\cal L}_{SB}$ is the symmetry breaking term of the QCD Lagrangian.
In the symmetric isospin limit it is given by \cite{GL,GSS} $ {\cal L}_{SB}= - \hat{m} \;\qb q$, where $q$ is the SU(2) quark field and $\hat{m}$ is the average quark mass: $\hm= (m_u+m_d)/2$.
At leading order in the chiral expansion the effective and fundamental symmetry breaking parameters are related by a constant, denoted by $B$: $\m^2 = 2B\;\hm$.
As $\hm$ and $\mu^2$ are small, we have 

\beq
\s_d = \hm \; \frac{dM}{d\hm} = \m^2 \; \frac{d M}{d \m^2} \;
\label{B2}
\eeq

\ni
and write $\s_d = 2\s_N + \s_\e$, where $\s_N=\m^2\;dm/d\m^2$ and  $\s_\e$ describes the changes in the condensate 
as compared to an assembly of static non-interacting nucleons.

In the framework of the Schr\"odinger equation, the binding energy is

\beq
- \e = \int d^3 r\;\psi^{*} \lp - \frac{\nabla^2}{m} + V \rp \psi \;,
\label{B3}
\eeq

\ni
where $ \psi $ is the deuteron wave function. 
Thus

\bea
-\;\frac{d\e}{d\m^2} &=& \int d^3 r \lb \psi^* \lp\frac{\s_N}{\m^2}\;  \frac{\nabla^2}{m^2}
+ \frac{d V}{d \m^2}\rp \psi 
+ \frac{d\psi^*}{d\m^2}\lp - \frac{\nabla^2}{m} + V \rp \psi
+ \psi \lp - \frac{\nabla^2}{m} + V \rp \frac {d\psi^*}{d\m^2}\rb
\nn\\[2mm]
&=& \int d^3\br \lb \psi^* \lp \frac{\s_N}{\m^2}\;  \frac{\nabla^2}{m^2}+ \frac{d V}{d \m^2}\rp \psi 
-\e \; \frac{d }{d \mu^2}\lp \psi^* \psi \rp  \rb  \;.
\label{B4}
\eea

The term proportional to $\e$ in this result does not contribute when the deuteron wave function is kept properly normalized and
we write

\beq
\s_\e = \int d^3 r \;  \psi^* \lp \s_N \; \frac{\nabla^2}{m^2}+\m^2 \; \frac{dV}{d\m^2}\rp \psi \;.
\label{B5}
\eeq

\ni
The first term on the r.h.s. of this equation is the effect of the scalar nucleon number and reduces the sigma commutator by 
a factor $(1-T/m)$, where $T$ is the nucleon kinetic energy, as compared to the additive assumption.
Using the equation of motion, we have

\beq
\s_\e = \int d^3 r \;  \psi^* \lb \frac{\s_N}{m}(V+\e) +\m^2 \; \frac{dV}{d\m^2}\rb \psi \;.
\label{B6}
\eeq

The contribution proportional to $\e$ is tiny and will not be considered in the sequence.
The deuteron is heavily dominated by the one pion exchange potential ($V_\pi$) and we write the full NN interaction as 

\beq
V = \Vb_{\pi}  + W \;,
\label{B7}
\eeq

\ni
where $\Vb_\p$ is the OPEP regularized at small distances and $ W $ represents other short and medium range effects, associated with either meson or quark dynamics.
In the absence of a theory for the influence of chiral symmetry breaking over both $W$ and the regularizing potential, we assume that these functions do not depend explicitly on $\m$. 

For the deuteron channel one has $\btu^{(1)}\!\cdot\!\btu^{(2)} = -3 $ and the OPEP reads

\beq
V_\p = -\lp \frac{g_A}{f_\p}\rp^2 \frac{\m^3}{16 \p} 
\lb \bsg^{(1)}\!\cdot\! \bsg^{(2)}\lp  U_C - G \rp + S_{12}\; U_T \rb \;,
\label{B8}
\eeq
\ni
where

\bea
U_C &=& \frac{e^{-\m r}}{\m r} ,
\label{B9}\\[2mm]
U_T &=& \lp 1 + \frac{3}{\m r} + \frac{3}{\m^2 r^2} \rp \frac{e^{-\m r}}{\m r} 
\eea
\label{B10}

\ni
and G is proportional to a delta-function: $ G = 4 \pi / \mu^3 \; \delta^3(r) $. 
The effects of this last term are cancelled by the regularization procedure and we skip them in the sequence.

The derivative of $V_\p$ with respect to $\m^2$ is 

\bea
\frac{d V_\p}{ d \m^2}
&=&  2\; \frac{f_\p}{g_A} \lp \frac{d}{d\m^2}\;\frac{g_A}{f_\p}\rp V_\p
+ \frac{1}{2} \lp \frac{g_A}{f_{\pi}}\rp^2  \frac{\mu}{16 \pi}
\lb \bsg^{(1)}\!\cdot\! \bsg^{(2)} \lp  1\! - \frac{2}{\m r} \rp  + S_{12} \lp 1\! + \frac{1}{\m r} \rp \rb e^{-\m r}
\nn\\[2mm]
&\equiv&  2\; \frac{f_\p}{g_A} \lp \frac{d}{d\m^2}\;\frac{g_A}{f_\p}\rp V_\p
+ \lp \frac{d V_\p}{ d \m^2}\rp_{\frac{g_A}{f_\p}} \;.
\label{B11}
\eea

This allows eq.(\ref{B6}) to be written as  

\beq
\s_\e = \la \m^2 \; \frac{d\Vb_\p}{d\m^2} \ra_{\frac{g_A}{f_\p}}
+c\; \la \Vb_\p \ra  \;.
\label{B12}
\eeq

\ni
with

\bea
&& \la \Vb_\p \ra  \equiv  \int d^3 r \; \psi^* \; \Vb_\p \;  \psi \;,
\label{B13}\\[2mm]
&& \la \m^2 \frac{d\Vb_\p}{d\m^2} \ra_{\frac{g_A}{f_\p}}
\equiv   \int d^3 r \; \psi^* \;\m^2\; \lp \frac{d\Vb_\p}{d\m^2} \rp_{\frac{g_A}{f_\p}}\psi \;,
\label{B14}
\eea

\ni
and

\beq
c = \frac{\s_N}{m}+  2\m^2 \lp \frac{1}{g_A}\; \frac{d g_A}{d\m^2}-\; \frac{1}{f_\p}\;\frac{d f_\p}{d \m^2} \rp  \;. 
\label{B15}
\eeq

The quantity $\s_\e$ represents the part of the deuteron $\s$-term due to NN intraction and may be probed by scalar sources.
In practice, these sources may be associated with either photons or pions, as we discuss in the next sections.
In order to interpret eq.(\ref{B12}), one notes that the coefficient $c$, given by eq.(\ref{B15}), vanishes in the chiral limit:
$\m^2=0 \Rightarrow c=0$.
Hence, at tree level, only the first term contributes, which represents the interaction of the scalar source with the 
pion exchanged between the nucleons.
The coefficient $c$, on the other hand, receives contributions from the kinetic energy term and from the derivative of the 
$\p$NN coupling constant. 
The latter, as we show in the sequence, corresponds to the interaction of the scalar source with the pion cloud that
dresses the $\p$N vextex.

In order to estimate the derivative of $f_\pi$, we use the result produced by Gasser and Leutwyler  \cite{GL} and write:

\bea
\frac{d f_\p}{d\m^2} 
&=& \frac{d}{d\m^2}\lc F \lb 1+\frac{\m^2}{F^2}\lp \ell_4^r(\l) 
-\frac{1}{16 \p^2} \ln \frac{\m^2}{\l^2}\rp\rb\rc 
\nn\\[2mm]
&=& \frac{1}{F} \lb \ell_4^r(\l) - \frac{1}{16\p^2} \lp 1+ \ln {\frac{\m^2}{\l^2}}\rp \rb  \;,
\label{B16}
\eea

\ni
where $F$ is the value of $f_\p$ for $\m=0$, $\ell_4^r(\l)$ is a renormalization constant and $\l$ is the renormalization 
scale. As far as the derivative of $g_A$ is concerned, we  use the expression derived by Moj\v{z}i\v{s}\cite{M} and by
Fearing, Lewis, Mobed and Scherer \cite{FLMS} and have

\bea
\frac{d g_A}{d\m^2} 
&=& \frac{d}{d\m^2} \lc G_A \lb 1+ \frac{4\m^2}{m^2}a_3 - \frac{\m^2 G_A^2}{16\p^2 F^2}
-\frac{\m^2}{16\p^2F^2}\lp 1+2 G_A^2 \rp \ln \frac{\m^2}{\l^2}\rb + \frac{\m^2}{4\p^2 F^2} b_{17}^r(\l) \rc
\nn\\[2mm]
&=&  G_A \lb \frac{4 a_3}{m^2} - \frac{G_A^2}{16\p^2 F^2}
-\frac{1}{16\p^2F^2}\lp 1+2 G_A^2 \rp \lp1+ \ln \frac{\m^2}{\l^2}\rp \rb + \frac{1}{4\p^2 F^2} b_{17}^r(\l)  \;,
\label{B17}
\eea

\ni
where $G_A$ is the value of $g_A$ in the limit $\m\rar 0$ and $b_{17}^r(\l)$ is a constant. 
Note that the expression adopted for $g_A$, within curly brackets,  is slightly different from that obtained earlier by Bernard, Kaiser and Meissner
\cite{BKM} and consistent \cite{UM} with that produced by Gasser, Sainio and \v{S}varc\cite{GSS}.

For future purposes, we write down the following results

\bea
\la V_\p \ra &=& - \lp \frac{ g_A}{f_\p}\rp^2  \frac{\m^3}{16 \pi} \int dr \lb \; u^2
\right.
\nn\\[2mm]
&+& \left.
2 \sqrt{8} \lp 1+\frac{3}{\m r}+\frac{3}{\m^2 r^2}\rp u w - \lp 1+\frac{6}{\m r}+\frac{6}{\m^2 r^2}\rp w^2\; \rb
\frac{e^{-\m r}}{\m r} \;,
\label{B18}\\[2mm]
\la \m^2\; \frac{d V_\p}{d\m^2} \ra_{\frac{g_A}{f_\p}}
&=&  \lp \frac{ g_A}{f_\p}\rp^2  \frac{\m^3}{32\pi} \int dr \lb \lp \m r - 2 \rp u^2
\right.
\nn\\[2mm]
&+& \left. 2 \sqrt{8} \lp \m r + 1\rp u w - \lp \m r + 4 \rp w^2 \; \rb  \frac{e^{-\m r}}{\m r}\;,
\label{B19}
\eea

\ni 
where $u$ and $w$ are the standard S and D components of the deuteron wave function. 
These expressions contain negative powers of r, but this does not pose problems for the integration, even in the case of 
unregularized potentials, since $ u $ and $ w $ vanish at the origin.
The numerical implications of the results presented here will be explored in sect. V.
We now discuss some possible ways of probing the many-body effects of the condensate.


\section{electromagnetic probes}

A probe which couples locally to the pion field $\bfi $ is sensitive to the quantity $ \la A| \bfi^2 | A \ra $, i.e., to the nuclear condensate.
In particular, when a nucleus A is probed by electromagnetic interactions, the many body effects of the condensate correspond to meson exchange contributions to the forward Compton amplitude $ F^A_{mec}(0)$, for soft photons. 
This relationship was established by Chanfray and Ericson \cite{CE93}, using the static approximation, but it is more general and its derivation does not require this assumption. 
Indeed,  in their work on the extension of the Bethe-Levinger sum rule, Ericson, Rosa-Clot and Kulagin \cite{ERK96} have shown that $ F^A_{mec}(0) $ contains a pion exchange term, which is the seagull represented in fig. 1(a) and can be expressed as:

\beq
F^A_{mec}(0) = - \frac{2}{3} \; e^2 \; \int d\br \lp \la A | \bfi^2 | A \ra -A \la N | \bfi^2 | N \ra \rp  \;,
\label{C1}
\;.
\label{C2}
\eeq

\ni
The second term in the r.h.s. of eq.(\ref{C1}) represents the expectation value of $\bfi^2$ for an assembly of free nucleons,
which has to be subtracted to obtain the exchange piece.
On the other hand, the matrix element $\la A | \bfi^2 | A \ra $ is related to the quark condensate by ${\cal L}_{SB}$,  the chiral symmetry breaking term in the Lagrangian for the $SU(2)$ sector, as discussed by Chanfray and Ericson \cite{CE93}.
In the case of QCD one has ${\cal L}_{SB} = - \hm \;\bar{q}q $, assuming $ m_u = m_d = \hm $. 
This symmetry breaking term transforms according to the $ (\frac{1}{2},\frac{1}{2}) $ representation of $ SU(2) \times SU(2) $ and one requires the same to happen with the effective counterpart.
In the case of non-linear realizations of the symmetry, this corresponds to the choice

\beq
{\cal L}_{SB} = \m^2 f_\p \sqrt{f_\p^2-\bfi^2}\;.
\label{C3}
\eeq

\noindent
Imposing the equivalence of the fundamental and effective descriptions, we obtain

\bea
\la A | {\cal{L}}_{SB} | A \ra &=&
-  \hm\; \la A | \bar{q}q  | A \ra 
\nn\\
&=&  \m^2 f_\p^2 -\frac{1}{2} \m^2 \la A | \bfi^2 | A \ra +\cdots
\label{C4}
\eea

\ni
In the case of the vacuum, it yields the Gell-Mann-Oakes and Renner relation:
$ -\hm\;\la 0 | \bar{q}q | 0 \ra  = \m^2 f_\p^2 $. 
We apply this relation to both nuclei and free nucleons.
Using these results in eq.(\ref{C1}), we obtain the following relation between the condensate and the meson exchange Compton amplitude

\beq
F_A^{exch}(0) = -\;\frac{4}{3 }\;e^2 \; f_\pi^2
\int d\br \lp \frac{\la A | \bar{q}q | A \ra- A \la N | \bar{q}q | N \ra}{\la 0 | \bar{q}q | 0 \ra}\rp \;,
\label{C5}
\eeq

\ni
which is the same result of ref.\cite{CE93}, but now obtained without the use of the static approximation.
In the case of the deuteron, the exchange amplitude is related to the $\s_\e$ calculated in the previous section through

\beq
F_A^{exch}(0) = \frac{4\;e^2}{3\;\m^2}\s_\e 
\;.
\label{C6}
\eeq

Two comments on formula (\ref{C5}) are in order.
The soft photon amplitude on deuteron is given by the Thomson limit: $F_d(0)= -e^2/M$.
The exchange part $F_A^{exch}(0)$ is hidden in this term together with other contributions and they all add up to the
Thomson value.
The second remark concerns the composition of $\s_\e$, built of three terms: the kinetic energy term, the derivative of the 
$\p$NN coupling constant, and the derivative of the pion propagator.
When transposed into the Compton amplitude, the third part gives rise to the usual meson exchange term of fig. 1(a), where
two photons interact with an exchanged pion.
The derivative of the $\p NN$ coupling attaches the two photons to the $\p NN$ vertex, fig.1(b).
As far as the kinetic energy term is concerned, the fact that $\bfi^2$ is a scalar object means that its expectaton value involves
a $\bar{\psi}\psi$ combination of nucleon fields, which displays the same reduction factor $(1-T/m)$ as
the sigma commutator, with respect to the ordinary nucleon density.
Similar remarks apply to pion rescattering.
Numerical values will be discussed in Sect.V.

\section{pion probes}

Pions exchanged between nucleons may also be probed by means of external pions.
In this section we consider $a _{mec}$, the MEC contribution to the pion-deuteron scattering lenght.
The quadri-momenta for pions at rest are $k = k' =\lp \o ,\bz \rp $, where $ \o =\m$ or $0$ depending on whether the pions 
are physical or soft.
The $ \p d $ scattering length is generically given by

\beq
a\lp\o \rp = \frac{\m}{2\pi \lp 1+ \m / M \rp}
\int d\br \; \psi^*\lp\br  \rp  t\lp\br;\omega\rp \psi\lp\br\rp \;,
\label{D1}
\eeq

\ni
where $t$ is the part of  the amplitude for the process $ \pi N N \rightarrow \pi N N $ which does not contain two positive 
energy  nucleons propagating forward in time.

When PCAC holds, the sigma commutator is related to the soft pion scattering amplitude.
Hence the value of $\s_\e$ is associated with many body effects in the soft pion PCAC amplitude, since
$a_{mec}^{PCAC}(0)\;\a\; \s_\e$ \cite{ER72}.
We confront this relation with the direct evaluation of $a_{mec}(\mu)$, the quantity accessible to experiment.
The structure of this amplitude was already discussed in ref.\cite{R80} and here we are interested in its relationship with $\s_\e$.
This question is important because it concerns the possibility of obtaining empirical information about $\s_\e$ from measurements of the $\p$d scattering length.

For soft pions, the operator $t_{mec}$ is completely dominated by processes involving only pions and nucleons, whereas
for physical pions there are other contributions, mainly due to $ \Delta $ excitations.

In the $\p$N sector,  the basic interactions are obtained from the following non-linear Lagrangian, approximately 
invariant under SU(2)$\times$SU(2)
 
\bea
{\cal{L}}^{int}_{\p N} &=& 
\frac{1}{8 f_{\p}^2}\lb  \ds \bfi^2 \di \bfi^2 - \m^2 \bfi^4 \rb 
+ \frac{g_A}{2 f_\p} \bar{N} \g^\m\g_5\btu N\! \cdot\! \di \bfi
\nonumber\\
&-&\frac{1}{4f_\p^2} \bar{N} \g^\m \btu N \!\cdot\! \bfi \times\di \bfi
+ \frac{g_A}{8 f_\p^3} \bar{N} \g^\m\g_5\btu N\! \cdot\! \bfi\;\di \bfi^2
+\cdots\;,
\label{D3} 
\eea

\ni
designed to be used in the tree approximation.

The meson exchange currents are given by the diagrams shown in fig.2, which contain pion propagators coupled to nucleons.
Hence it is useful to parametrize the non relativistic MEC contribution to $ t $ in the nucleon sector as

\beq
t_{mec}^N(\bq; \o ) = \frac{1}{2\m}  \lp \frac{g_A}{2f_\p}\rp^2
\lc\lb \sum \a_n (\o) \rb
\frac{\bsg^{(1)}\!\cdot \!\bq \; \bsg^{(2)}\!\cdot\!\bq}{(\bq^2+\m^2)} 
+ \a'_1 (\o)\;\m^2\; \frac{\bsg^{(1)}\!\cdot\!\bq \; \bsg^{(2)}\!\cdot\!\bq}{(\bq^2+\mu^2)^2}\rc \;,
\label{D4}
\eeq

\noindent
where $\bq$ is the momentum exchanged between the nucleons and the coefficients $\alpha_n$ are determined dynamically, from the graphs of fig.2. 

The evaluation of the diagrams 1-10 of fig.2 in the non relativistic tree approximation yields \cite{R80}

\bea
&&\a_1 = \frac{2}{f_\p^2} ,
\label{Da1}\\
&&\a'_1 =  \frac{1}{f_\p^2} \lp 3-2\;\frac{\o^2}{\m^2}\rp,
\label{Da2}\\
&&\a_2 = -\; \frac{2}{f_\p^2} ,
\label{Da3}\\
&&\a_3 + \a_4 + \a_5 + \a_6 = \frac{2}{f_\p^2}\;\frac{\o^2}{m^2},
\label{Da4}\\
&&\a_7 + \a_8 = \lp \frac{g_A}{2f_\p}\rp^2 \frac{\o^2}{m^2} ,
\label{Da5}\\
&&\a_9+\a_{10} = 0 .
\label{Da6}
\end{eqnarray}

\noindent
As discussed in ref.\cite{RW78}, there is a cancellation between $\alpha_1$ and $\alpha_2$, required by chiral symmetry.
The results for $ \alpha_3+\alpha_4 + \alpha_5+\alpha_6 $ and $ \alpha_7+\alpha_8 $ disagree with those of ref.\cite{R80} by  factors ($-1$) and (-$\frac{3}{2}$) respectively, due to algebraic mistakes in that work, but this has little influence over
numerical results. 

The MEC amplitude in configuration space is

\beq
t_{mec}^N (\br; \o ) = \frac{1}{2\m}\; \frac{1}{3}\;
\lc \lb \sum \a_n (\o) \rb \ V_\p(r)
- \a'_1 (\o) \;\m^2\; \lp \frac{dV_\p(r)}{d\m^2}\rp_{\frac{g_A}{f_\p}} \rc \; .
\end{equation}
\label{D5}

We now consider the contributions of the $\Delta$ and $\s_N$ to $t_{mec}$. 
The former  were studied in ref.\cite{R80} and its efect can be incorporated into eq.(\ref{D4}) by means of the global coefficient $\a_\D \;\o^2/\m^2$, with $\a_\D = -0.429 \m^{-2}$.
The contribution of the $\pi$N sigma-term is given by diagrams 1-4 of fig.3 and can be calculated by noting that it 
enters only in the isospin symmetric $\pi$N amplitude $A^+$. 
The corresponding part of this amplitude is denoted by $A_\s^+$ and can be parametrized as \cite{TM79}

\beq
A_\s^+ \lp t;k^2,k'^2 \rp = \frac{\s_N}{\m^2 f_\p^2}
\lb k'^2+k^2 - \m^2 + \b \lp t-k'^2-k^2 \rp \rb 
\label{D6}
\eeq

\ni
and the value of $ \b $ can be extracted from scattering data.
The evaluation of the diagrams of fig.3 yields, for the coefficients $\a$,

\beq
\a_{\s 1} +\a_{\s 2} + \a_{\s 3} +\a_{\s 4} = 
\frac{4}{m} \frac{\s_N}{f_\p^2 \m^2} \lb \o^2- \lp  \bq^2+\m^2  \rp\rb \;.
\label{D7}
\eeq

The term proportional to $\lp \bq^2+\m^2 \rp$ cancels the pion propagator, giving rise to a contact interaction, which does not 
contribute when the OPEP is regularized.
The overall MEC contribution to the scattering length then becomes

\bea
a_{mec}(\o) &=& \frac{1}{4\p (1+ \m / M)}\; \frac{1}{3 f_\p^2}
\lc \lb \lp 2+\frac{g_A^2}{4}\rp \frac{\o^2}{m^2}
+f_\p^2 \;\a_\D\;\frac{\o^2}{\m^2} +\frac{4 \s_N }{m}\;\frac{\o^2}{\m^2}\rb 
\la V_\p \ra
\right.
\nn\\[2mm]
&-& \left.
\lp 3 - 2\;\frac{\o^2}{\m^2} \rp \;
\la \m^2 \frac{dV_\p}{d\m^2}\ra_{\frac{g_A}{f_\p}}\rc 
\label{D8}
\eea

In order to establish the relationship between $a_{mec}(\o)$ and $\s_\e$, we use eq.(\ref{B12}) and write

\bea
a_{mec}(\o) &=& \frac{1}{4\p (1+ \m / M)}\; \frac{1}{3 f_\p^2}
\lc \lb \lp 2+\frac{g_A^2}{4}\rp \frac{\o^2}{m^2}
+f_\p^2 \;\a_\D\;\frac{\o^2}{\m^2} + \frac{4 \s_N }{m}\;\frac{\o^2}{\m^2}
\right.\right.
\nn\\[2mm]
&+& \left.\left. c\; \lp 3 - 2\;\frac{\o^2}{\m^2} \rp \rb \la V_\p \ra
- \lp 3 - 2\;\frac{\o^2}{\m^2} \rp \; \s_\e \rc \;.
\label{D9}
\eea

In the soft pion limit $(\o\rar 0)$ this result becomes 

\beq
a_{mec}(0) = -\;\frac{1}{4\p (1+ \m / M)}\; \frac{\s_\e-c\;\la V_\p \ra}{ f_\p^2} \;.
\label{D10}
\eeq

For physical pions, on the other hand, one has 

\beq
a_{mec}(\m) = -\; \frac{1}{4\p (1+ \m / M)}\; \frac{1}{3 f_\p^2}
\lc\s_\e - \lb \lp 2+\frac{g_A^2}{4}\rp \frac{\m^2}{m^2}
+ f_\p^2 \;\a_\D  + \frac{4 \s_N }{m}+c \rb
\la V_\p \ra \rc \;.
\label{D11}
\eeq

The first observation from eq.(\ref{D10}) is that $a_{mec}(0)$ is not just proportional to $\s_\e$, as in the PCAC result,
$a_{mec}^{PCAC}(0)$, but the term $c\; \la V_\p \ra$ which appears in the epression (\ref{B12}) of $\s_\e$ is
cancelled in $a_{mec}(0)$.
The reason for this difference is that the usual meson exchange amplitude, $a_{mec}$, does not incorporate terms where 
the two pions are attached to the $\p NN$ vertex through loop diagrams.
These terms are instead present in the PCAC expression.
The fact that the term in $c\; \la V_\p \ra$ may give a large contribution to $\s_\e$ indicates a possible importance also as 
an exchange correction.
Moreover, inspecting eqs.(\ref{D10}) and (\ref{D11}), one notes that the contribution proportional to $dV_\p/d\m^2$  is three times larger for soft pions than for physical
pions, due to the strong energy dependence of the intermediate $ \p\p $ amplitude of diagram 1.
This feature is consistent with the results found by Chanfray, Ericson and Wambach \cite{CEW96}, who studied the self energy 
$\Pi(\o,\bk)$ of a pion propagating in a gas of of  pions. 
Using PCAC and the Hartree approximation, they found that

\beq
\Pi(\o,\bk) = \frac{\r_s}{f_\p^2}
\lb \m^2 - \frac{2}{3} \lp \o^2-\bk^2\rp\rb \;,
\label{D12}
\eeq

\ni
where $\r_s$ is the scalar density of the pions.
Thus, for soft and physical pions one has, respectively, $\Pi(0,0)= \r_s \m^2/f_\p^2$ and
$\Pi(\m,0)= \r_s \m^2/3f_\p^2$. 
As this self energy is related to the MEC amplitude, both must change in the same proportion when one goes from physical to 
soft pions.

In summary, the measurable meson exchange contribution written in eq.(\ref{D11}) has little relation to the quark condensate. 
Therefore, the pion-deuteron scattering length provides no exploitable information about this condensate. 
In the next section we discuss numerically the results produced here.


\section{results and conclusions}

We estimate the numerical implications of the results produced in the previous sections and adopt the following values for the
various constants: M=1875.61 MeV\cite{PDG}, m=938.28 MeV \cite{H}, $\m$=139.57 MeV \cite{H}, 
g$_A$=1.26 \cite{H}, f$_\p$=93.3 MeV \cite{H}, $\s_N$=45 MeV \cite{GLS91}, $\a_\D$=-0.43 $\m^{-2}$ \cite{R80},
$\l=\m$\cite{GL}, $\ell_4^r(\m) = 4.3/16\p^2$\cite{GL}, and $a_3=- m \s_N/4\m^2$\cite{M}.
As very little is known about the constant $b_{17}^r(\m)$, we neglect it in eq.(\ref{B17}).
With these inputs, we find a negative value for $c$: -0.30, which is strongly dominated by the derivative of the $\p$N 
coupling constant and has opposite sign to the kinetic energy term. 
Thus one has \\

$\lb \lp 2+g_A^2/4\rp \m^2/m^2
+f_\p^2 \;\a_\D + 4 \s_N /m  + c\; \rb
= \lb 0.05-0.19+0.19 -0.30 \rb=-0.25\;$ .\\

Expressions (\ref{B18}) and (\ref{B19}) are based on the assumption that the short range components of the interaction are 
not important since the OPEP strongly dominates the deuteron. 
In order to test this hypothesis, we consider the case of a toy potential containing an OPEP tail and regularized by means of monopole form factors \cite{BR94}.
It has the same form as eq.(\ref{B8}), with $ U_C $,  $ G $ and $ U_T $ given by

\bea
U_C &=& \frac{e^{-\m r}}{\m r}-\;\frac{\L_C}{\m} \frac{e^{-\L_C r}}{\L_C r} 
-\frac{1}{2}\; \frac{\m}{\L_C} \lp \frac{\L_C^2}{\m^2} -1 \rp e^{-\L_C r} \;,
\label{E1}\\[2mm]
G &=& \d \; \frac{1}{2}\; \frac{\m}{\L_C} \lp \frac{\L_C^2}{\m^2} -1 \rp^2 e^{-\L_C r} \;,
\label{E2}\\[2mm]
U_T &=& \lp 1+\frac{3}{\m r}+\frac{3}{\m^2 r^2} \rp \frac{e^{-\m r}}{\m r}
-\;\frac{\L_T^3}{\mu^3}\lp 1+\frac{3}{\L_T r}+\frac{3}{\L_T^2 r^2} \rp \frac{e^{-\L_T r}}{\L_T r} 
\nn\\[2mm]
&-& \frac{1}{2}\; \frac{\L_T}{\m}\lp \frac{\L_T^2}{\m^2}-1 \rp \lp 1+\L_T r \rp \frac{e^{-\L_T r}}{\L_T r} \;,
\label{E3}
\eea

\ni
where $\L_C$ and $\L_T$ are cut-offs for the central and tensor components and the parameter $\d$ regulates the strength of the short range function $ G $. 
The pure OPEP results are recoverd in the limit $ \L_C = \L_T \rar \infty $ and $ \d = 1$.
It yields a regularized version of eqs.(\ref{B18}) and (\ref{B19}), namely 

\bea
&& \la V_\p \ra = -\lp\frac{g_A}{f_\p}\rp^2 \frac{\m^3}{16\p}\int dr 
\lb \lp U_C\!-\!G \rp u^2 +2\sqrt{8}\; U_T \;uw  +\lp U_C\!-\!G\!-\!2U_T \rp w^2 \rb \;,
\label{E4}\\[2mm]
&&\la\m^2 \frac{dV_\p}{d\m^2} \ra_{\frac{g_A}{f_\p}} = \frac{3}{2}\; \la V_\p \ra
\nn\\[2mm]
&& -\lp\frac{g_A}{f_\p}\rp^2 \frac{\m^3}{16\p}\int dr \;\m^2
\lb \frac{d \lp U_C\!-\!G\rp }{d\m^2}\; u^2 
+2\sqrt{8}\; \frac{d U_T }{d\m^2} \;uw  + \frac{d \lp U_C\!-\!G\!-\!2U_T \rp}{d\m^2}\; w^2 \rb \;,
\label{E5}
\eea

\ni
with 

\bea
\m^2 \; \frac{d U_C}{d \m^2} &=& -\;\frac{1}{2} 
\lb \lp 1+\frac{1}{\m r} \rp \; e^{-\m r}
-\;\frac{e^{-\L_C r}}{\m r} 
-\;\frac{1}{2}\; \frac{\m}{\L_C} \lp \frac{\L_C^2}{\m^2} + 1 \rp e^{-\L_C r} \rb \;,
\label{E6}\\[2mm]
\m^2\; \frac{d G}{d \m^2} &=& - \d \; \frac{1}{4}\; \frac{\m}{\L_C} 
\lp 3 \frac{\L_C^4}{\m^4} - 2 \frac{\L_C^2}{\m^2} -1 \rp e^{-\L_C r} \;,
\label{E7}\\[2mm]
\m^2\;\frac{d U_T}{d \m^2} &=& \;-\frac{1}{2} \lb \lp 1+\frac{4}{\m r}+\frac{9}{\m^2 r^2} +\frac{9}{\m^3 r^3} \rp e^{-\m r}
- 3 \;\frac{\L_T^3}{\mu^3}\lp 1+\frac{3}{\L_T r}+\frac{3}{\L_T^2 r^2} \rp \frac{e^{-\L_T r}}{\L_T r} 
\right.
\nn\\[2mm]
&-& \left. \frac{1}{2}\; \frac{\L_T}{\m}\lp 3\;\frac{\L_T^2}{\m^2}-1 \rp \lp 1+\L_T r \rp \frac{e^{-\L_T r}}{\L_T r} \rb \;,
\label{E8}
\eea

In general, the deuteron binding energy is a function of the form $\e (g_A,f_\p,\m,\L_C,\d,\L_T ) $. 
As $g_A$, $f_\p$ and $\m$ are kept fixed, the binding energy  depends on the the short range parameters $ \L_C$, $ \d$ and 
$ \L_T $ .  
When constructing the deuteron, we fix two of them and look for the third one so as to have  $ \epsilon = 2.2250  $ MeV.

\begin{table}[hbt]
\caption{Deuteron expectation for $V_\p$ and $\m^2 dV_\p/d\m^2$, for the perturbative (OPEP) and
regularized (toy) one-pion exchange potentials, eqs.(\ref{B18}, \ref{B19}) and eqs.(\ref{E4}, \ref{E5}),
$\s_\e$, eq.(\ref{B12}), $a_{mec}(0)$, eq.(\ref{D10}), 
$a_{mec}(\m)$, eq.(\ref{D11}) as functions of the inner parameters $\d$, $\L_C$ and $\L_T$. 
The values quoted for $\L_C$ and $\L_T$ were rounded up.}
\begin{tabular} {|c|c|c||c|c||c|c|c|c|c|}

&&& OPEP & OPEP & toy & toy & toy & toy & toy\\

$\d$
 & $\L_{C}$ 
 & $\L_{T}$
 & $\la V_\p \ra $
 & $\la \m^2 \frac{d V_\p}{d \m^2} \ra $  
 & $\la V_\p \ra $
 & $\la \m^2 \frac{d V_\p}{d \m^2} \ra $ 
 & $\s_\e$
 & $a_{mec} (0)$
 & $a_{mec} (\m)$\\

& (GeV) & (GeV) & (MeV) & (MeV) & (MeV) & (MeV) & (MeV)  & ($\m^{-1}$) & ($\m^{-1}$)\\
\hline \hline

 1 & 1.579 & 1.086 & -60.94 & 1.93 & -25.71 & 1.73 & 9.44  & -0.0021 & -0.0012 \\ \hline
 1 & 1.973 & 1.054 & -58.39 & 1.91 & -25.72 & 1.70 & 9.42  & -0.0020 & -0.0012 \\ \hline
 1 & 2.368 & 1.028 & -57.93 & 1.89 & -25.73 & 1.67 & 9.39 &  -0.0020 & -0.0012 \\ \hline
 1 & 2.763 & 1.008 & -58.30 & 1.87 & -25.72 & 1.63 & 9.35 & -0.0019 & -0.0011 \\ \hline
 1 & 3.157 & 0.992 & -59.00 & 1.85 & -25.69 & 1.60 & 9.31 & -0.0019 & -0.0011 \\ \hline \hline

 5 & 1.579 & 1.809 & -49,65 & 2.15 & -29.21 & 1.96 & 10.72 & -0.0023 & -0.0013 \\ \hline
 5 & 1.973 & 1.457 & -45.76 & 2.09 & -28.54 & 1.97 & 10.53 & -0.0023 & -0.0013 \\ \hline
 5 & 2.368 & 1.288 & -46.01 & 2.05 & -28.14 & 1.94 & 10.38 & -0.0023 & -0.0013 \\ \hline
 5 & 2.763 & 1.194 & -47.24 & 2.01 & -27.83 & 1.89 & 10.24 & -0.0022 & -0.0013 \\ \hline
 5 & 3.157 & 1.134 & -48.74 & 1.99 & -27.57 & 1.84 & 10.11 & -0.0022 & -0.0013 \\ \hline \hline

10 & 1.973 & 2.223 & -43.88 & 2.14 & -29.99 & 2.03 & 11.03 & -0.0024 & -0.0014 \\ \hline
10 & 2.368 & 1.558 & -42.67 & 2.09 & -29.35 & 2.02 & 10.83 & -0.0024 & -0.0014 \\ \hline
10 & 2.763 & 1.344 & -43.74 & 2.06 & -28.85 & 1.97 & 10.63 & -0.0023 & -0.0013 \\ \hline
10 & 3.157 & 1.233 & -45.29 & 2.03 & -28.46 & 1.92 & 10.46 & -0.0023 & -0.0013 \\ 
\end{tabular}
\label{Tab.1}
\end{table}

In table 1 we display our results for $ \la V_\p \ra$ and $\la \m^2 \;d V_\p /d \m^2 \ra $ as given by the 
the perturbative OPEP (pert) eqs. (\ref{E4}) and  (\ref{E5}) and by the regularized OPEP (toy), eqs.(\ref{B18}) mand (\ref{B19}). 
The first feature to be noted is that the sensitivity to the regularization of the potential is much 
greater for $\la V_\p \ra$ than for $ \la \m^2 \; d V_\p/d\m^2 \ra $, due to the fact that the latter is 
less influenced by the short distance components of the wave function.
In the case of the calculation based on the regularized potential, the large variations of the inner parameters considerd change results only by a few percent.
This suggests that the self consistency between the potential and the wave function is important.
In table 2 we present our results for the case of the Argonne $v_{14}$  \cite{A84} and super soft core C \cite{S73} potentials and the values quoted also follow the pattern found in the case of the toy potential.

\begin{table}[hbt]
\caption{ Deuteron expectation for $V_\p$ and $\m^2 dV_\p/d\m^2$, for the perturbative (OPEP) and
regularized one-pion exchange potentials, eqs.(\ref{B18}, \ref{B19}) and eqs.(\ref{E4}, \ref{E5}),
$\s_\e$, eq.(\ref{B12}), $a_{mec}(0)$, eq.(\ref{D10}), 
$a_{mec}(\m)$, eq.(\ref{D11})
and $F_A^{exch}(0)$, eq.(\ref{C6}),  the MEC contribution to the electromagnetic form factor,
for the Argonne 
and SSC  
realistic interactions.}
\begin{tabular} {|c||c|c||c|c|c|c|c|c|}

& OPEP & OPEP & & & & & & \\

potential   
 & $\la V_\p \ra $
 & $\la \m^2 \frac{d V_\p}{d \m^2} \ra $  
 & $\la V_\p \ra $
 & $\la \m^2 \frac{d V_\p}{d \m^2} \ra $ 
 & $\s_\e$
 & $a_{mec} (0)$
 & $a_{mec} (\m)$
 & $F_A^{exch}(0)$\\

&(MeV)&(MeV)& (MeV) & (MeV) & (MeV) & ($\m^{-1}$) & ($\m^{-1}$)&($e^2\m^{-1}$) \\
\hline \hline

Argonne & -33.63 & 1.80 & -19.83 & 1.52 & 7.47 & -0.0019 & -0.0011 & 0.071\\ \hline
SSC       & -29.27 & 1.57 & -14.94 & 1.48 & 5.96 & -0.0019 & -0.0011 & 0.057 \\  
\end{tabular}
\label{Tab.2}
\end{table}

Inspection of these tables shows that the expectation values of the potential are about ten times larger than those of its derivative.
Taking this information into eq.(\ref{D4}), one finds that this corresponds to an average pion momentum $q=3\m$, which is relatively high.
The disturbance of the QCD vacuum due to the NN interaction, represented by $\s_\e$, has a central value of
about 10 MeV, which is about five times the binding energy  and corresponds to about 10\% of the total deuteron 
$\s$ term. 
Our results have the same magnitude but an opposite sign to that produced by Gammal and Frederico \cite{GF98} in the framework of  the Skyrme model.  
The values of $\s_\e$ quoted in the tables are dominated by the component involving the constant $c$ in eq.(\ref{B12}).
This in turn depends strongly on $d g_A/d\m^2$ which was calculated using chiral perturbation theory and contains an 
unknown constant.
Hence our result has to be taken as an estimate of the magnitude of $\s_\e$.

The columns $a_{mec}(0)$, eq.(\ref{D10}) and $a_{mec}(\m)$, eq.(\ref{D11}), correspond respectively to  the quantities that have a relation to the condensate $\s_\e$.
The difference between $a_{mec}(0)$ and $a_{mec}(\m)$ stems in part from the factor 3, related to the off-shell 
behaviour of the intermediate pion-pion scattering amplitude, as discussed at the end of section IV.
In the case of soft pions, it is worth noting that $\frac{1}{3}  a_{mec}(0) \approx -0.0007 \mu^{-1}$, in 
agreement with the value found by Robilotta and Wilkin for physical pions \cite{RW78}.
The value for $F_A^{exch}(0)$, the many body electromagnetic term of the commutator amplitude, is also displayed.

In summary, we have studied the many body effects of the quark condensate in the deuteron through the Feynman-Hellmann theorem and found out that the part of the deuteron sigma commutator associated with the NN interaction is smaller than the pion-nucleon sigma-term, but five times larger than the binding energy.
With the restricions mentioned previously ($b_{17}^r$ is not known), we find that $\s_\e$ could be dominated by the derivative 
of the $\p$N coupling constant.
We have also linked the changes in the condensate with meson exchange effects for probes that can couple to the pion field, namely Compton and pion scatterings. 
As far as the possibility of extracting  $\s_\e$ from the pion-deuteron scattering length, our study has shown that
meson exchange effects are comparable to the present experimental error \cite{EX00}. 
However the extrapolation to the soft limit produces important changes which tend to blur the contribution of $\s_\e$.
The reason why the pion-deuteron scattering length is unexploitable is that the part of the exchange correction which is linked
to the sigma commutator is reduced by a factor 3 when one goes from soft to physical pions, which makes it small. 
Moreover, in the last case, non static corrections appear, in such a way that the extraction of the interesting term becomes 
unfeasible.
In the case of the Compton amplitude, instead, no such problem arises, since soft photons are directly accessible to experiment,
opening the the possibility of empirical determination. 
The photons are by far a superior tool as a source of information on the quark condensate, not only in the deuteron, but also in 
nuclei.

\vs{10mm}

\vspace{10mm} 
{Acknowledgments}
 
We would like to thank G. Chanfray,  J. Delorme,  C.A. Dominguez and  H. Leutwyler for useful discussions,
 J. Gasser,  J. Goity and M. Moj\v{z}i\v{s} for exchanges of messages, and U-G. Meissner for help in dealing with aspects
chiral perturbation theory.
It is also our pleasure to acknowledge the hospitality of the Institute of Nuclear Theory and the Nuclear Theory Group of the University of Washington, USA, where this work was initiated. 
M.R.R. would also like to thank the hospitality of the Division de Physique Theorique de l'Institut de Physique Nucleaire,
Orsay, France and FAPESP, for financial support.

{}


\vspace{2cm}
{\large \textbf{Figure Captions}} \vspace{2cm}

\noindent
Fig.1 Seagull meson exchange diagram contributing to the Compton amplitude.

\noindent
Fig.2. Diagrams contributing to the pion-deuteron scattering length in the pure pion-nucleon sector;
the crosses in the propagators of figs. 9 and 10 indicate that they refer to antinucleons.

\noindent
Fig.3. Diagrams contributing to the pion-deuteron scattering length
due to the isospin-symmetric amplitude represented by the black square.

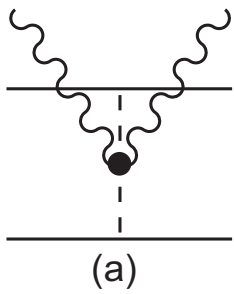
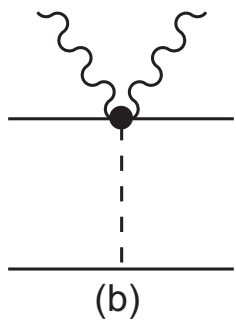
\end{document}